\documentclass[a4paper,11pt]{article}
\usepackage{jheppub}
\usepackage[T1]{fontenc} 
\usepackage{pdfsync}
\usepackage{hyperref}
\usepackage{multirow}
\usepackage{color}
\usepackage{slashed}
\usepackage{subfig}
\usepackage[utf8x]{inputenc}

\title{\boldmath Off-shell chromomagnetic dipole moments in the SM at and beyond the $Z$ gauge boson mass scale}

\author[a,b]{J. Monta\~no-Dom\'inguez,}
\author[a]{B. Quezadas-Vivian,}
\author[a,1]{F. Ram\'irez-Zavaleta,\note{Corresponding author.}}
\author[a]{E. S. Tututi}
\author[a]{and E. Urquiza-Trejo}

\affiliation[a]{Facultad de Ciencias F\'isico Matem\'aticas, Universidad Michoacana de San Nicol\'as de Hidalgo, Av. Francisco J. M\'ugica s/n, C.~P. 58060, Morelia, Michoac\'an, M\'exico}
\affiliation[b]{C\'atedras Conacyt, Av. Insurgentes Sur 1582, Col. Cr\'edito Constructor, Alc. Benito Ju\'arez, C.~P. 03940, Ciudad de M\'exico, M\'exico}

\emailAdd{jmontano@conacyt.mx}
\emailAdd{brenda.quezadas@umich.mx}
\emailAdd{feramirez@umich.mx}
\emailAdd{eduardo.tututi@umich.mx}
\emailAdd{everardo.urquiza@umich.mx}

\abstract{The off-shell anomalous chromomagnetic dipole moment of the standard model quarks ($u$, $d$, $s$, $c$ and $b$), at the $Z$ gauge boson mass scale, is computed by using the $\overline{\textrm{MS}}$ scheme. The numerical results disagree with all the previous predictions reported in the literature and show a discrepancy of up to two orders of magnitude in certain situations.}

\begin{document}
\maketitle
\flushbottom

\section{Introduction}
\label{sec:intro}

Recently, an experimental measurement of an anomalous chromomagnetic dipole moment (CMDM) has been reported at CERN. The CMS Collaboration \cite{Sirunyan:2019eyu,PDG2020} announced the top quark CMDM measurement
\begin{eqnarray}
\hat{\mu}_t^\mathrm{Exp}=-0.024_{-0.009}^{+0.013}(\mathrm{stat})_{-0.011}^{+0.016}(\mathrm{syst}),\nonumber
\end{eqnarray}
and the chromoelectric dipole moment (CEDM) bound $|\hat{d}_t^\mathrm{~Exp}|<0.03$, at $95\%$ C.~L. Theoretically, the top quark off-shell CMDM prediction in the Standard Model (SM) was reported in detail in Refs.~\cite{Aranda:2018zis,Aranda:2020tox}, where $\hat{\mu}_t(-m_Z^2)$ $=$ $-0.0224$$-$$0.000925i$ and
$\hat{\mu}_t(m_Z^2)$ $=$ $-0.0133$$-$$0.0267i$. In Ref.~\cite{Aranda:2020tox} the infrared (IR) divergence in the CMDM was carefully analyzed, which arises from the non-Abelian triple gluon vertex diagram when the gluon is on-shell ($q^2=0$). In previous works, it has been demonstrated that the off-shell CMDM is consistent with an observable since it is gauge invariant, gauge independent, IR finite and ultraviolet (UV) finite~\cite{Aranda:2020tox,Davydychev:2000rt,Hernandez-Juarez:2020drn}. These properties were corroborated in the non-linear $R_{\xi}$ gauge~\cite{Aranda:2020tox} and using the background field method~\cite{Hernandez-Juarez:2020drn}. Therefore, for consistency, in this work we also computed this observable taking the gluon off-shell ($q^2\neq 0$) at large-momentum-transfer, as it occurs with the perturbative strong running coupling constant $\alpha_s(m_Z^2)$~\cite{Deur:2016tte}. For this reason, we evaluated the off-shell CMDM in two scenarios: the spacelike ($q^2=-m_Z^2$) and the timelike one ($q^2=m_Z^2$). Accordingly, in this work we perform analogous evaluations for the quarks $u$, $d$, $s$ (light), $c$ and $b$.

In the perturbative regime, the SM CMDM ($\hat{\mu}_q$) is induced at the one-loop level by quantum chromodynamics (QCD) and electroweak (EW) contributions. In the SM, the CMDM of quarks with the off-shell gluon has been studied in Refs.~\cite{Hernandez-Juarez:2020drn} and \cite{Choudhury:2014lna}. In Ref.~\cite{Choudhury:2014lna} the off-shell CMDM of the SM quarks was studied at the spacelike value $q^2=-m_Z^2$, nevertheless, we have identified issues that have been clarified in Ref.~\cite{Aranda:2020tox}, therefore we recalculate such scenario for the $u$, $d$, $s$, $c$ and $b$ quarks. On the other hand, Ref.~\cite{Hernandez-Juarez:2020drn} only reports the timelike evaluation $q^2=m_Z^2$, however, the running of the quark masses at the scale of the $Z$ gauge boson mass is not considered by the authors. In contrast, our work takes into account both scenarios ($q^2=\pm m_Z^2$), where we employ the running of the quark masses in the $\overline{\textrm{MS}}$ scheme at the $Z$ gauge boson mass scale, which leads to important corrections.

The outline of this paper is as follows. In Sec.~\ref{Sec:CMDM-lagrangian} the general effective chromoelectromagnetic dipole moment Lagrangian is presented. In Sec.~\ref{Sec:CMDM-diagrams} the six one-loop diagram contributions to the off-shell CMDM are calculated. In Sec.~\ref{Sec:results} the numerical results of the off-shell CMDM for each quark are discussed. Sec.~\ref{Sec:conclusions} is devoted to our conclusions. In Appendix~\ref{appendix-Input-values} the input values are listed, where the quark masses are evaluated according to the $\overline{\textrm{MS}}$ scheme at the scale of the $Z$ gauge boson mass. The Appendix~\ref{appendix-PaVe} contains some Passarino-Veltman functions that were used.

\section{The chromomagnetic dipole moment}
\label{Sec:CMDM-lagrangian}

\begin{figure}[h!]
\centering
\includegraphics[width=7.0cm]{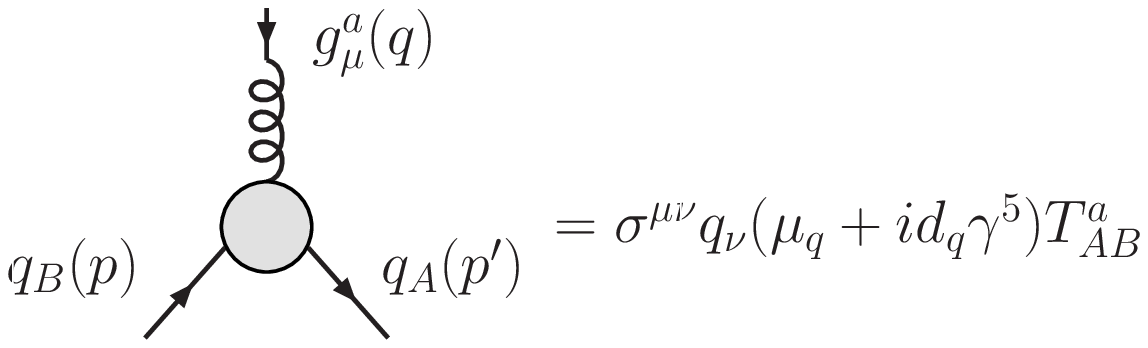}
\caption{Chromoelectromagnetic dipole moments.}
\label{FIGURE-CEMDM-general}
\end{figure}

The CMDM vertex is given in Fig.~\ref{FIGURE-CEMDM-general}, where $\mu_q$ represents the chromomagnetic dipole moment for any SM quark and $d_q$ the chromoelectric one (CEDM); more details about the corresponding quark-antiquark-gluon interaction Lagrangian is given in Ref.~\cite{Aranda:2020tox}. In the SM the CMDM is induced as a quantum fluctuation at the one-loop level
\cite{Aranda:2020tox,Choudhury:2014lna}, while the CEDM does not arise at this level. Here, the dimensionless dipole definitons are written as follows \cite{Khachatryan:2016xws,PDG2020,Bernreuther:2013aga,Haberl:1995ek}
\begin{equation}\label{}
\hat{\mu}_{q}\equiv \frac{m_q}{g_s}\mu_q \quad , \quad \hat{d}_q\equiv \frac{m_q}{g_s}d_q ~,
\end{equation}
where $m_q$ is the quark mass, $g_s=\sqrt{4\pi\alpha_s}$ is the QCD group coupling constant, being $\alpha_s$ the perturbative strong coupling constant characterized at the $Z$ gauge boson mass scale, with $\alpha_s(m_Z^2)=0.1179$ \cite{PDG2020}.

\section{The off-shell CMDM in the SM at the one-loop level}
\label{Sec:CMDM-diagrams}

In the SM the off-shell CMDM of a quark is composed by the sum of six contributions
\begin{eqnarray}\label{CMDM-complete}
\hat{\mu}_{q_i}(q^2)&=&\hat{\mu}_{q_i}(\gamma)+\hat{\mu}_{q_i}(Z)+\hat{\mu}_{q_i}(W)+\hat{\mu}_{q_i}(H)+\hat{\mu}_{q_i}(g)+\hat{\mu}_{q_i}(3g)~;
\end{eqnarray}
they are illustrated in Fig.~2 from Ref.~\cite{Aranda:2020tox}. In the following, we refer to them as
(a) $\hat{\mu}_{q_i}(\gamma)$ the Schwinger-type photon,
(b) $\hat{\mu}_{q_i}(Z)$ the $Z$ neutral gauge boson,
(c) $\hat{\mu}_{q_i}(W)$ the $W$ charged gauge boson,
(d) $\hat{\mu}_{q_i}(H)$ the Higgs boson,
(e) $\hat{\mu}_{q_i}(g)$ the Schwinger-type gluon,
and (f) $\hat{\mu}_{q_i}(3g)$ the triple gluon vertex.
The EW contribution is the sum of the (a)-(d) contributions and the QCD one is the sum of (e) and (f) diagrams.

The analytical result for each off-shell CMDM contribution has already been presented in Ref. \cite{Aranda:2020tox}; most of their analytical Passarino-Veltman functions (PaVe) are valid for generic quark masses, that is to say, if the mass is small (for $u$, $d$, $s$, $c$, $b$) or heavy (as the top quark). Below, we briefly comment on each contribution, indicating, when necessary, whether it should be considered an analytical distinction between light and heavy quarks.

\begin{figure}[h!]
\centering
\includegraphics[width=14.0cm]{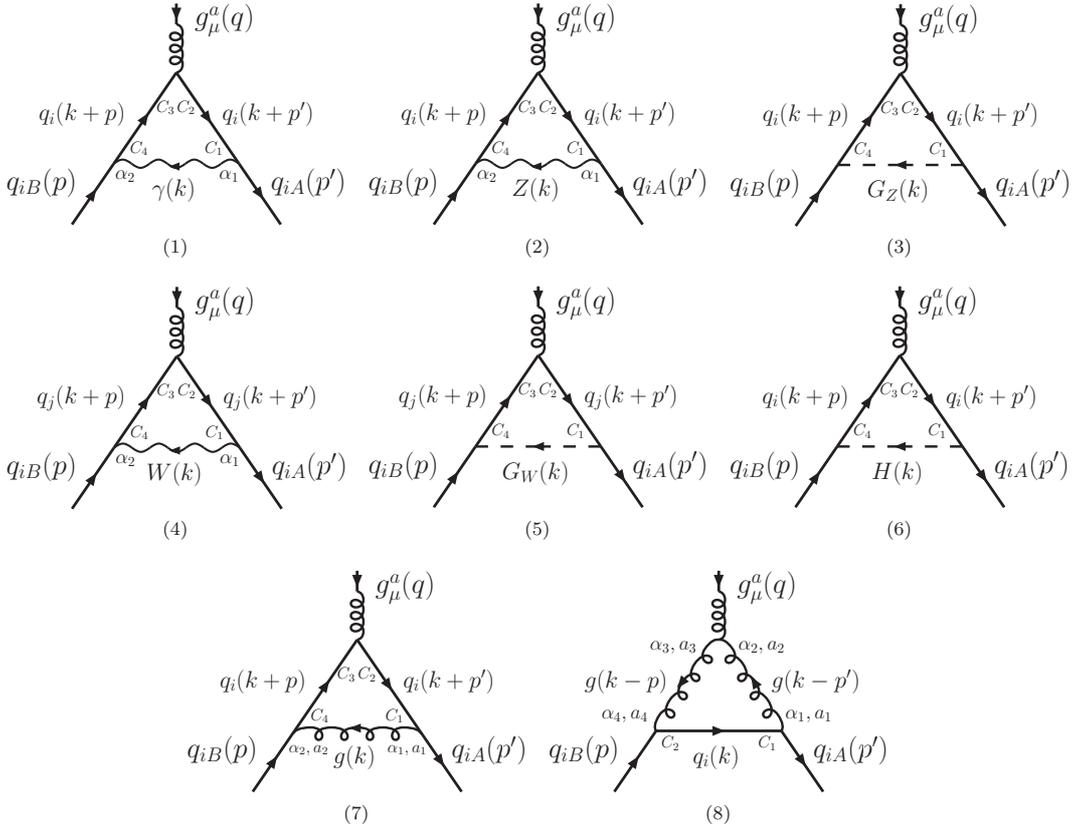}
\caption{Feynman diagrams that contribute to the chromoelectromagnetic dipole moments in the covariant $R_\xi$ gauge.}
\label{CCMDM-rxigauge}
\end{figure}
It is important to note that $\hat{\mu}_{q_i}(q^2)$ is compatible with the properties of an observable, since its calculation does not depend on the gauge parameter; all contributions to $\hat{\mu}_{q_i}(q^2)$ have been computed in the covariant $R_\xi$ gauge (see Fig.~\ref{CCMDM-rxigauge}), by means of the Passarino-Veltman reduction scheme. On the other hand, by using the Feynman parameterization method along with the covariant $R_\xi$ gauge, we explicitly demonstrate that the associated amplitude to the triple gluon vertex does not depend on the gauge parameter. Similarly, this procedure can be applied, without difficulty, for each of the contributions where the SM gauge fields and the pseudo-Goldstone bosons circulate in the loops (see Fig.~\ref{CCMDM-rxigauge}); we have taken the task of showing this result in order to provide an explanatory example.

\subsection{The $\gamma$ contribution}

The corresponding off-shell CMDM ($\hat{\mu}_{q_i}(\gamma)$) is given in Eq.~(11) from Ref.~\cite{Aranda:2020tox}. For this case, as it was shown in a previous work~\cite{Aranda:2020tox} (see Eq. (11)), it is found that $\hat{\mu}_{q_i}(\gamma)\propto m_{q_i}^2$, which leads to suppressed values for light quarks arises. The numerical evaluations for each SM quark are listed in Tables~\ref{TABLE-chromo-up}-\ref{TABLE-chromo-bottom}.

\

\subsection{The $Z$ gauge boson contribution}

The respective off-shell CMDM ($\hat{\mu}_{q_i}(Z)$) is given in Eq.~(14) of the Ref.~\cite{Aranda:2020tox}. In this  case, because $m_Z>m_{q_i}$, the scalar PaVe $C_0^Z$ $=$ $C_0\left(m_q^2,m_q^2,q^2,m_q^2,m_Z^2,m_q^2\right)$ function must be used (see Appendix~\ref{appendix-PaVe}). The numerical values for each SM quark are presented in Tables~\ref{TABLE-chromo-up}-\ref{TABLE-chromo-bottom}.

\

\subsection{The $W$ gauge boson contribution}

For the $\hat{\mu}_{q_i}(W)$ contribution, two diagram configurations must be distinguished. First, when an up-type quark is on-shell (see Fig.~2(c) in Ref.~\cite{Aranda:2020tox}), the $W^-$ boson and the down-type quarks are inside the loop. Second, when the down-type quark is on-shell, the $W^+$ boson and the up-type quarks are within the loop.

The off-shell CMDM for $u_i=u,c$ quarks contains the following components
\begin{equation}\label{}
\hat{\mu}_{u_i}(W)=\sum_{j=1}^3\hat{\mu}_{u_i}(W,d_j),
\end{equation}
where $d_j$ $=$ $d_1,d_2,d_3$ $=$ $d,s,b$~\cite{Aranda:2020tox}. Conversely, the off-shell CMDM for $d_i$ $=$ $d,s,b$ quarks is made-up by the sum
\begin{equation}\label{}
\hat{\mu}_{d_i}(W)=\sum_{j=1}^3\hat{\mu}_{d_i}(W,u_i),
\end{equation}
where $u_j$ $=$ $u_1,u_2,u_3$ $=$ $u,c,t$~\cite{Aranda:2020tox}. The corresponding general off-shell CMDM is given in Eq.~(16) of the Ref.~\cite{Aranda:2020tox}.

For $\hat{\mu}_{u,c}(W,d_j)$, it is true that $m_Z>m_W>m_{q_i,q_j}$, whilst
for the $\hat{\mu}_{d_i}(W,u)$ and $\hat{\mu}_{d_i}(W,c)$ contributions $m_Z>m_W>m_{q_i,q_j}$ is satisfied. In all previous contributions the scalar PaVe $C_0^W$ $=$ $C_0(m_{q_i}^2,m_{q_i}^2,q^2,m_{q_j}^2,m_W^2,m_{q_j}^2)$ must be used; relevant analytical expressions for this PaVe are given in Appendix~\ref{appendix-PaVe}. For $\hat{\mu}_{d_i}(W,t)$, where $m_{t}>m_Z>m_W>m_{q_i}$, the corresponding $C_0^W$ function employed can be found in Eq. (35) from Ref. \cite{Aranda:2020tox}. In Tables~\ref{TABLE-chromo-up}-\ref{TABLE-chromo-bottom} the numerical results for all SM quarks are listed.

\

\subsection{The Higgs boson contribution}

The generic off-shell CMDM for $\hat{\mu}_{q_i}(H)$ is given in Eq. (18) from Ref.~\cite{Aranda:2020tox}. In this situation, we have that $m_H>m_Z>m_{q_i}$, where the $C_0^H$ $=$ $C_0\left(m_q^2,m_q^2,q^2,m_q^2,m_H^2,m_q^2\right)$ must be used (see Eq. (\ref{C0-2}) in Appendix~\ref{appendix-PaVe}). The resulting values for each SM quark are presented in Tables~\ref{TABLE-chromo-up}-\ref{TABLE-chromo-bottom}.

\

\subsection{The $g$ contribution}

The off-shell CMDM contribution for $\hat{\mu}_{q_i}(g)$ can be viewed in Eq. (20) from Ref.~\cite{Aranda:2020tox}. This is analogous to the photon case, and it is evident that $\hat{\mu}_{q_i}(g)$ $\propto$ $m_{q_i}^2$, which provides suppressed values for small-mass quarks. The numerical evaluations for each SM quark are displayed in Tables~\ref{TABLE-chromo-up}-\ref{TABLE-chromo-bottom}.

\

\subsection{The $3g$ contribution}

In Eq. (23) from~\cite{Aranda:2020tox}, the off-shell CMDM for $\hat{\mu}_{q_i}(3g)$ can be found. For this part, all the analytical results presented in the mentioned Eq. (23) are valid for any SM quark. The numerical evaluation for all SM quarks are shown in Tables~\ref{TABLE-chromo-up}-\ref{TABLE-chromo-bottom}.

In the covariant $R_\xi$ gauge, the off-shell CMDM contribution of the triple gluon diagram, when $q^2\neq 0$, can be expressed as
\begin{eqnarray}\label{3g-integral-Rxi}
\mathcal{M}_{q_i}^\mu(3g) &=& \int\frac{d^Dk}{(2\pi)^D}\bar{u}(p')\left(-ig_s\gamma^{\alpha_1}T_{AC_1}^{a_1}\right)
\left(i\frac{\slashed{k}+m_{q_i}}{k^2-m_{q_i}^2+i\varepsilon}\delta_{C_1C_2}\right)\left(-ig_s\gamma^{\alpha_4}T_{C_2B}^{a_4}\right)u(p)
\nonumber\\
&&\times
\left\{\frac{i}{(k-p')^2+i\varepsilon}
\left[-g_{\alpha_1\alpha_2}+
(1-\xi)
\frac{(k-p')_{\alpha_1}(k-p')_{\alpha_2}}{(k-p')^2+i\varepsilon}\right]
\delta_{a_1a_2}\right\}
\nonumber\\
&&\times
\left[-g_sf_{aa_3a_2}T^{\mu\alpha_3\alpha_2}_{ggg}(p'-p,-k+p,k-p')\right]
\nonumber\\
&&\times
\left\{\frac{i}{(k-p)^2+i\varepsilon}
\left[-g_{\alpha_3\alpha_4}+(1-\xi)
\frac{(k-p)_{\alpha_3}(k-p)_{\alpha_4}}{(k-p)^2+i\varepsilon}\right]
\delta_{a_3a_4}\right\},
\end{eqnarray}
where $T_{AC_1}^{a_1}\delta_{C_1C_2}T_{C_2B}^{a_4}\delta_{a_1a_2}f_{aa_3a_2}\delta_{a_3a_4}$$=$
$-i\frac{3}{2}T_{AB}^a$ and $T_{\mu\alpha_3\alpha_2}^{ggg}(p'-p,-k+p,k-p')$ $\equiv$
$(k-2p+p')_{\alpha_2}g_{\mu\alpha_3}+(-2k+p+p')_\mu g_{\alpha_3\alpha_2}+(k+p-2p')_{\alpha_3}g_{\alpha_2\mu}$. By solving these integrals employing the Feynman parameterization and after some algebraic manipulations, it is found that
\begin{eqnarray}
\hat{\mu}_{q_i}(3g)_{R\xi} &=& \frac{\alpha_s}{32\pi}\int_0^1dx_1\int_0^{1-x_1}dx_2
~24(1-x_1-x_2)x_2\bigg[
\frac{m_q^2}{\Delta}4 (5 x_1-3) x_1
\nonumber\\
&&
+ \frac{m_q^2}{\Delta^2} \Big\{ 2 m_q^2 x_1^3 (4-5 x_1)
+q^2 x_1 \left[x_1^2 (5-10 x_2)-2 x_1 (x_2-1) (5 x_2-4)\right.\nonumber\\
&&\left.+8 (x_2-1) x_2+3\right]
\Big\}+ \frac{m_q^2}{\Delta^3} \Big(
2 m_q^4 (x_1-1) x_1^5
+q^2 \left\{2 m_q^2 (x_1-1) x_1^3 [x_1 (2 x_2-1)\right.\nonumber\\
&&\left.+2 (x_2-1) x_2+1]
\right\}
+2 (q^2)^2 (x_1-1) x_1 (x_2-1) x_2 (x_1+x_2-1) (x_1+x_2)
\Big)\nonumber\\
&&+(1-\xi) f_1(x_1,x_2)
+(1-\xi)^2 f_2(x_1,x_2)
\bigg],
\end{eqnarray}
where

\begin{eqnarray}\label{}
\Delta&\equiv& m_q^2 x_1^2+q^2 x_2 (x_1+x_2-1),\nonumber\\
f_1(x_1,x_2) &\equiv&
\frac{2 m_q^2 (5 x_1-2) x_1}{\Delta}
-\frac{1}{4 \Delta^2}\left[m_q^2 x_1 \left\{4 m_q^2 (5 x_1-6) x_1^2+q^2 \left[5 x_1^2 (4 x_2-1)\right.\right.\right.\nonumber\\
&&\left.\left.\left.+x_1 \left(20 x_2^2-24 x_2+7\right)-4 x_2^2+4 x_2-2\right]
\right\}\right]+\frac{1}{2 \Delta^3}\left\{m_q^2 x_1^2  \left[m_q^2 (x_1-2) x_1\right.\right.\nonumber\\
&&\left.\left.+q^2 x_2 (x_1+x_2-1)\right]
\left\{2 m_q^2 x_1^2+q^2 \left[x_1 (2 x_2-1)+2 x_2^2-2 x_2+1\right]
\right\}\right\},
\nonumber\\
f_2(x_1,x_2) &\equiv& \frac{m_q^2 q^2 x_1 (2 x_1-1)}{2 \Delta^2}
-\frac{m_q^2 q^2 x_1^2\left[m_q^2 (x_1-2) x_1+q^2 x_2 (x_1+x_2-1)\right]}{2 \Delta^3}.
\end{eqnarray}
The integrals proportional to the linear and quadratic gauge parameters are exactly zero:
\begin{eqnarray}\label{}
(1-\xi)\int_0^1dx_1\int_0^{1-x_1}dx_2~24(1-x_1-x_2)x_2f_1(x_1,x_2) &=& 0,
\nonumber\\
(1-\xi)^2\int_0^1dx_1\int_0^{1-x_1}dx_2~24(1-x_1-x_2)x_2f_2(x_1,x_2) &=& 0.
\end{eqnarray}
Therefore, it has been demonstrated that this specific contribution to $\hat{\mu}_{q_i}(q^2)$ is gauge independent.

\section{Results}
\label{Sec:results}

In order to evaluate the off-shell CMDM for each quark at the conventional EW reference scale of the $Z$ gauge boson mass in a consistent way, we take into account the quark masses at such scale, $m_q(m_Z)$, $q=u,d,s,c,b$, by running them in the corresponding renormalization group equations of QCD, making use of the RunDec package \cite{Chetyrkin:2000yt,Herren:2017osy} (see Appendix \ref{appendix-Input-values}). Our phenomenological analysis, where $\hat{\mu}_{q}$ is evaluated at the $Z$ boson mass scale, is also supported by an experimental study where the bottom-quark mass is determined at the energy scale of the $Z$ gauge boson mass~\cite{DELPHI}. We also consider all the pertinent input values at such reference scale. The numerical evaluation can be carried out with the provided Passarino-Veltman analytical solutions, with \texttt{Package-X}~\cite{Patel:2015tea}, or entirely with \texttt{Collier}~\cite{Denner:2016kdg}.

\begin{table}[!t]
  \centering
\begin{tabular}{|c|c|c|}\hline
\multirow{2}{*}{$\hat{\mu}_u$} & \multicolumn{2}{c|}{$q^2\hspace{2.5cm}$} \\
\cline{2-3}
& $-m_Z^2$  & $m_Z^2$  \\
\hline
$\gamma$ & $2.24\times10^{-12}$  &  $-2.24\times10^{-12}+{3.14\times10^{-13}}i$ \\
$Z$      & $-9.31\times10^{-14}$ &  $-1.23\times10^{-13}-{5.93\times10^{-14}}i$ \\
$W$      & $-1.5\times10^{-13}$  &  $-1.96\times10^{-13}-{1.\times10^{-13}}i$   \\
$H$      & $2.92\times10^{-23}$  &  $1.83\times10^{-23}+{3.79\times10^{-23}}i$  \\
$g$      & $-1.28\times10^{-11}$ &  $1.28\times10^{-11}-{1.79\times10^{-12}}i$  \\
$3g$     & $-1.05\times10^{-10}$ &  $1.05\times10^{-10}-{1.61\times10^{-11}}i$  \\
Total    & $-1.15\times10^{-10}$ &  $1.15\times10^{-10}-1.77\times10^{-11}i$    \\
\hline
\end{tabular}
\caption{Off-shell CMDM of the up quark.}\label{TABLE-chromo-up}
\end{table}


\begin{table}[!t]
  \centering
\begin{tabular}{|c|c|c|}\hline
\multirow{2}{*}{$\hat{\mu}_d$} & \multicolumn{2}{c|}{$q^2\hspace{2.5cm}$} \\
  \cline{2-3}
& $-m_Z^2$ & $m_Z^2$  \\
\hline
$\gamma$ & $2.44\times10^{-12}$  & $-2.44\times10^{-12}+3.67\times10^{-13}i$ \\
$Z$      & $-4.10\times10^{-13}$ & $-5.40\times10^{-13}-2.51\times10^{-13}i$ \\
$W$      & $-6.98\times10^{-13}$ & $-9.15\times10^{-13}-4.67\times10^{-13}i$ \\
$H$      & $6.37\times10^{-22}$  & $4.01\times10^{-22}+8.28\times10^{-22}i$  \\
$g$      & $-5.56\times10^{-11}$ & $5.56\times10^{-11}-8.36\times10^{-12}i$  \\
$3g$     & $-4.52\times10^{-10}$ & $4.52\times10^{-10}-7.53\times10^{-11}i$  \\
Total    & $-5.07\times10^{-10}$ & $5.04\times10^{-10}-8.4\times10^{-11}i$   \\
\hline
\end{tabular}
\caption{Off-shell CMDM of the down quark. }\label{TABLE-chromo-down}
\end{table}

\begin{table}[!t]
  \centering
\begin{tabular}{|c|c|c|}\hline
\multirow{2}{*}{$\hat{\mu}_s$} & \multicolumn{2}{c|}{$q^2\hspace{2.5cm}$} \\
  \cline{2-3}
         & $-m_Z^2$              & $m_Z^2$  \\
\hline
$\gamma$ & $6.90\times10^{-10}$  & $-6.90\times10^{-10}+1.45\times10^{-10}i$ \\
$Z$      & $-1.62\times10^{-10}$ & $-2.14\times10^{-10}-9.96\times10^{-11}i$ \\
$W$      & $-2.85\times10^{-10}$ & $-3.73\times10^{-10}-1.9\times10^{-10}i$ \\
$H$      & $1.00\times10^{-16}$  & $6.30\times10^{-17}+1.30\times10^{-16}i$ \\
$g$      & $-1.57\times10^{-8}$  & $1.57\times10^{-8}-3.32\times10^{-9}i$ \\
$3g$     & $-1.23\times10^{-7}$  & $1.23\times10^{-7}-2.98\times10^{-8}i$ \\
Total    & $-1.38\times10^{-7}$  & $1.37\times10^{-7}-3.33\times10^{-8}i$    \\
\hline
\end{tabular}
\caption{Off-shell CMDM of the strange quark.}\label{TABLE-chromo-strange}
\end{table}

Before comparing our predictions with the corresponding ones reported in the literature, we first describe the general results. Our evaluations for the quarks $u$, $d$, $s$, $c$, and $b$ are listed in Tables~\ref{TABLE-chromo-up}-\ref{TABLE-chromo-bottom}; they are ordered according to the mass hierarchy $m_u$ $<$ $m_d$ $<$ $m_s$ $<$ $m_c$ $<$ $m_b$.
The spacelike scenario ($q^2=-m_Z^2$) only yields real values. On the other hand, the timelike prediction ($q^2=-m_Z^2$) gives rise to complex quantities. By comparing the $\mathrm{Re}\thinspace\hat{\mu}_{q_i}(q^2)$ at $q^2=-m_Z^2$ against the $q^2=m_Z^2$ case (see Table~\ref{TABLE-chromo-all}), both have the same order of magnitude but opposite sign. In almost all cases, for the timelike scenario, the $\mathrm{Im}\thinspace\hat{\mu}_{q_i}(q^2)$ parts are of the same order of magnitude as the $\mathrm{Re}\thinspace\hat{\mu}_{q_i}$ ones, but with opposite sign. In terms of orders of magnitude, they are
$|\hat{\mu}_u|\sim 10^{-10}$,
$|\hat{\mu}_d|\sim 10^{-10}$,
$|\hat{\mu}_s|\sim 10^{-7}$,
$|\hat{\mu}_c|\sim 10^{-5}$, and
$|\hat{\mu}_b|\sim 10^{-4}$.
In all these cases the largest contributions come from $\hat{\mu}_{q_i}(3g)$ and the smallest ones from $\hat{\mu}_{q_i}(H)$.

\begin{table}[!t]
  \centering
\begin{tabular}{|c|c|c|c|}\hline
\multirow{2}{*}{$\hat{\mu}_c$} & \multicolumn{2}{c|}{$q^2\hspace{2.5cm}$} \\
  \cline{2-3}
           & $-m_Z^2$           & $m_Z^2$  \\
\hline
$\gamma$ & $2.53\times10^{-7}$  & $-2.53\times10^{-7}+{7.95\times10^{-8}}i$  \\
$Z$      & $-2.36\times10^{-8}$ & $-3.11\times10^{-8}-{1.50\times10^{-8}}i$  \\
$W$      & $-3.89\times10^{-8}$ & $-5.1\times10^{-8}-{2.6\times10^{-8}}i$    \\
$H$      & $1.87\times10^{-12}$ & $1.18\times10^{-12}+{2.43\times10^{-12}}i$ \\
$g$      & $-1.44\times10^{-6}$ & $1.44\times10^{-6}-{4.53\times10^{-7}}i$   \\
$3g$     & $-1.04\times10^{-5}$ & $1.04\times10^{-5}-{4.07\times10^{-6}}i$   \\
Total    & $-1.16\times10^{-5}$ & $1.15\times10^{-5}-{4.49\times10^{-6}}i$   \\
\hline
\end{tabular}
\caption{Off-shell CMDM of the charm quark. }\label{TABLE-chromo-charm}
\end{table}

\begin{table}[!t]
  \centering
\begin{tabular}{|c|c|c|}\hline
\multirow{2}{*}{$\hat{\mu}_b$} & \multicolumn{2}{c|}{$q^2\hspace{2.5cm}$} \\
  \cline{2-3}
         & $-m_Z^2$             & $m_Z^2$  \\
\hline
$\gamma$ & $9.40\times10^{-7}$  & $-9.43\times10^{-7}+4.29\times10^{-7}i$  \\
$Z$      & $-4.78\times10^{-7}$ & $-6.31\times10^{-7}-2.93\times10^{-7}i$  \\
$W$      & $-8.07\times10^{-7}$ & $-8.54\times10^{-7}-9.27\times10^{-10}i$ \\
$H$      & $8.62\times10^{-10}$ & $5.53\times10^{-10}+1.13\times10^{-9}i$  \\
$g$      & $-2.14\times10^{-5}$ & $2.15\times10^{-5}-9.78\times10^{-6}i$   \\
$3g$     & $-1.40\times10^{-4}$ & $1.36\times10^{-4}-8.56\times10^{-5}i$   \\
Total    & $-1.61\times10^{-4}$ & $1.55\times10^{-4}-9.52\times10^{-5}i$   \\
\hline
\end{tabular}
\caption{Off-shell CMDM of the bottom quark. }\label{TABLE-chromo-bottom}
\end{table}

\newpage

\subsection{The spacelike evaluation $q^2=-m_Z^2$}

Firstly, let us point out some issues in Ref. \cite{Choudhury:2014lna}, where the calculation for the off-shell CMDM of the SM quarks at $q^2=-m_Z^2$ has some points that need to be clarified. For example, the color algebra in the triple gluon diagram is incorrect (see footnote 1 in Ref. \cite{Aranda:2020tox}); the authors did not report the imaginary part, which is induced by the $W$ gauge boson contribution~\cite{Aranda:2018zis,Aranda:2020tox}. Now, we proceed to compare our off-shell CMDM results for each quark (see Table~\ref{TABLE-chromo-all}) with those from Ref. \cite{Choudhury:2014lna}. To do that, we have organized their results in Table~\ref{TABLE-indians}, where we identify that $\hat{\mu}=\Delta\kappa/4$. The authors only reported EW contributions for each diagram, however, did not provide individual evaluations for the two participating diagrams in the QCD case. Instead, the authors plotted the total contributions ($\Delta\kappa$) as a function of a fictitious virtual gluon mass $m_g$, from which we take (in order to compare) values at $m_g=0$. It should be noted that their results remain essentially independent of $m_g$, which implies that whenever $|q^2|>0$ it would be no longer necessary to assume a fictitious mass. In Table~\ref{TABLE-chromo-all} our off-shell CMDM results for all the quarks, at $q^2=-m_Z^2$, are presented. Notice that all our predictions are negative.

\begin{table}[!t]
  \centering
\begin{tabular}{|c|c|c|}\hline
\multirow{2}{*}{$\hat{\mu}_q$} & \multicolumn{2}{c|}{$q^2\hspace{2.2cm}$} \\
\cline{2-3}
& $-m_Z^2$ & $m_Z^2$  \\
\hline
$\hat{\mu}_u$ & $-1.15\times10^{-10}$ & $1.15\times10^{-10}-1.77\times10^{-11}i$  \\
$\hat{\mu}_d$ & $-5.07\times10^{-10}$ & $5.04\times10^{-10}-8.4\times10^{-11}i$   \\
$\hat{\mu}_s$ & $-1.38\times10^{-7}$  & $1.37\times10^{-7}-3.33\times10^{-8}i$    \\
$\hat{\mu}_c$ & $-1.16\times10^{-5}$  & $1.15\times10^{-5}-{4.49\times10^{-6}}i$   \\
$\hat{\mu}_b$ & $-1.61\times10^{-4}$  & $1.55\times10^{-4}-9.52\times10^{-5}i$   \\
\hline
\end{tabular}
\caption{Total values for the off-shell CMDMs of small-mass quarks at the $Z$ gauge boson mass scale.}
\label{TABLE-chromo-all}
\end{table}

\begin{table}
  \centering
\begin{tabular}{|c|c|}\hline
$\hat{\mu}_{q_i}$  & Total \\
\hline
$\hat{\mu}_u$  & $2.15\times10^{-11}$   \\
$\hat{\mu}_d$  & $6.50\times10^{-11}$   \\
$\hat{\mu}_s$  & $3.5\times10^{-9}$     \\
$\hat{\mu}_c$  & $1.79\times10^{-6}$    \\
$\hat{\mu}_b$  & $-2.01\times10^{-5}$   \\
\hline
\end{tabular}
\caption{Off-shell CMDMs of small-mass quarks at $q^2=-m_Z^2$ and $m_g=0$ from Ref.~\cite{Choudhury:2014lna}, where $\hat{\mu}=\Delta\kappa/4$.}\label{TABLE-indians}
\end{table}

Comparing our results with those from Table~\ref{TABLE-indians}, it can be appreciated that the absolute value for $\hat{\mu}_u$, $\hat{\mu}_d$, $\hat{\mu}_c$, and $\hat{\mu}_b$ is one order of magnitude smaller, whilst $\hat{\mu}_s$ is two orders of magnitude smaller.

\begin{figure}[h!]
\centering
\includegraphics[width=9.0cm]{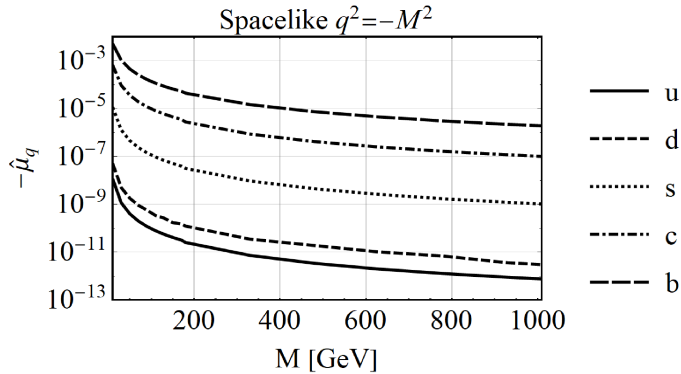}
\caption{CMDM of the SM quarks as a function of the gluon momentum transfer $q^2=-M^2$ (spacelike scenario), $M=[10,1000]$ GeV.}
\label{gsl}
\end{figure}
The next analysis considers the running of the quark masses from 10 GeV up to one TeV~\cite{Herren:2017osy}. Specifically, in Fig. \ref{gsl} we analyze the behavior of the off-shell CMDM at energies beyond the $Z$ boson mass scale, in the plot it can be appreciated the decoupling nature of this quantity. If the $Z$ boson mass is taken as a point of reference, it is observed that $\hat{\mu}_q$ decreases as the energy increases, decreasing its value around 2 orders of magnitude for all the quarks under consideration, with the exception of the up quark, where the gap is of 3 orders of magnitude. In this case, that corresponds to the spacelike scenario, in the energy range studied, between 10 and 1000 GeV, no imaginary contribution to $\hat{\mu}_q$ is induced.

\subsection{The timelike evaluation $q^2=m_Z^2$}

In the timelike scenario a recent paper \cite{Hernandez-Juarez:2020drn} address the study of the off-shell CMDM for the SM quarks. The authors used the Feynman parameterization method as in Ref. \cite{Choudhury:2014lna}, and found some other disagreements with this work. These authors reproduced our previous result for $\hat{\mu}_t(m_Z^2)$ \cite{Aranda:2018zis,Aranda:2020tox}, and mention that they also agree with our $\hat{\mu}_t(-m_Z^2)$ (see Sec. 3.1.2 in Ref. \cite{Hernandez-Juarez:2020drn}). In addition, they confirmed the fact that the $W$ gauge boson yields an imaginary part.

Nevertheless, the authors in Ref.~\cite{Hernandez-Juarez:2020drn} have an issue in their triple-gluon CMDM when the gluon is on-shell. Even though their CMDM in Eq.~(14) for the off-shell gluon agrees with our Eq.~(23)~\cite{Aranda:2020tox}, through $B_0(0,m_q^2,m_q^2)=B_0(m_q^2,0,m_q^2)-2$, they claim that their Eq.~(15) is IR divergent, which is incorrect (see Eq. (26) from Ref.~\cite{Aranda:2020tox}). Actually, they present an UV divergence, which is due to the fact that they incorrectly take $B_0(0,0,0)=0$ from \texttt{Package-X}, since this contains both UV and IR divergences. In the user manual of \texttt{Package-X} it is established that the software employs the same $1/\epsilon$ pole symbol to indicate an UV divergence or an IR one, that is why $B_0(0,0,0)$ falsely disappears.

Then, comparing our CMDMs for the timelike scenario ($q^2=m_Z^2$), listed in Table~\ref{TABLE-chromo-all}, with those from Table 1 in Ref.~\cite{Hernandez-Juarez:2020drn}, it can be observed that our evaluations are smaller. This is a consequence of considering the running of the quark masses at the scale of the $Z$ gauge boson mass, in the $\overline{\textrm{MS}}$ scheme. In specific, the absolute values of our CMDMs are smaller in:
$\hat{\mu}_u$ by $34\%$,
$\hat{\mu}_d$ by $35\%$,
$\hat{\mu}_s$ by $36\%$,
$\hat{\mu}_c$ by $28\%$,
and
$\hat{\mu}_b$ by $54\%$.

\begin{figure}[h!]
\centering
\subfloat[]{\includegraphics[width=7.5cm]{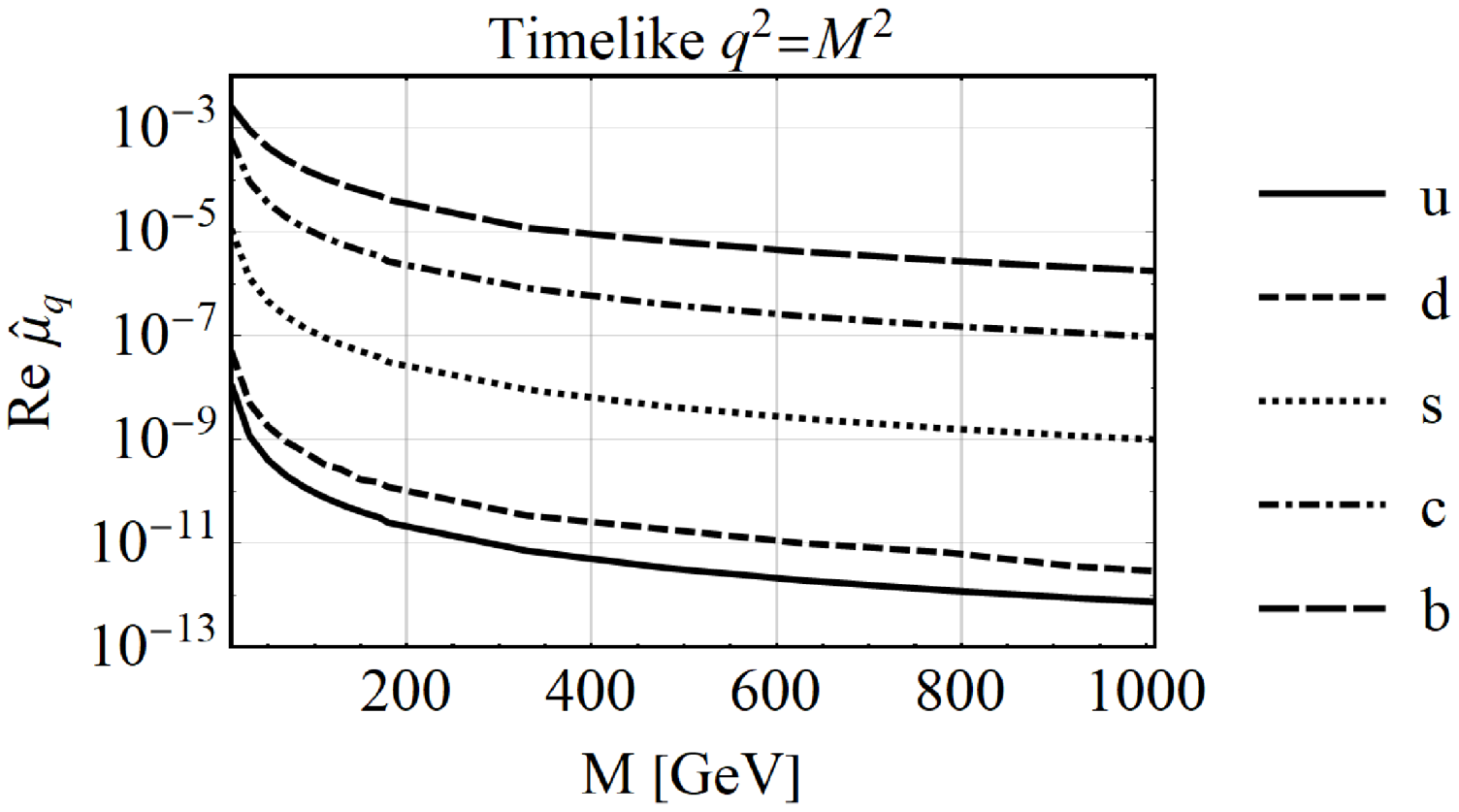}}
\subfloat[]{\includegraphics[width=7.5cm]{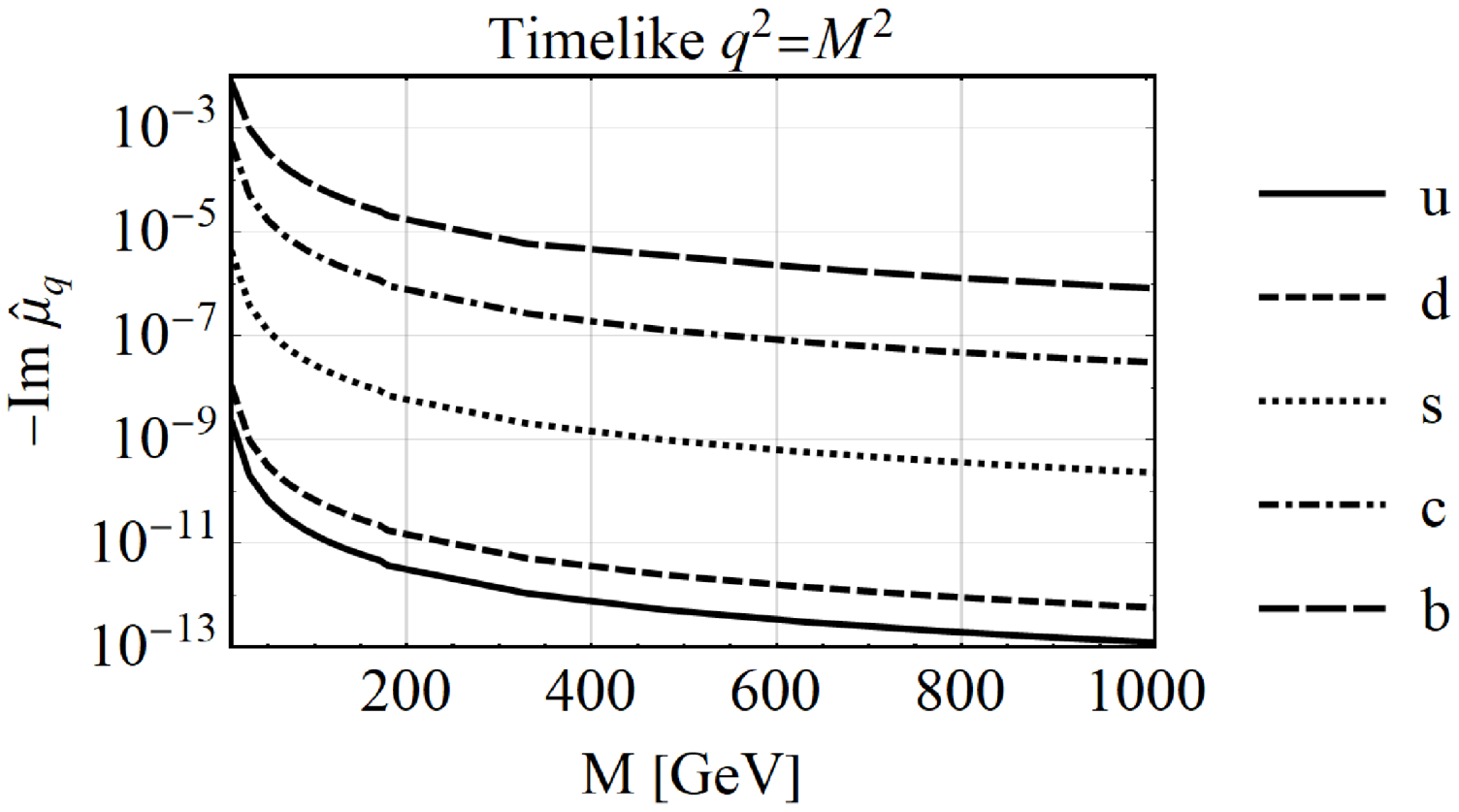}}
\caption{CMDM of the SM quarks as a function of the gluon momentum transfer $q^2=M^2$ (timelike scenario), $M=[10,1000]$ GeV.}
\label{gtl}
\end{figure}
In Fig.~\ref{gtl} the off-shell CMDM at energies between 10 and 1000 GeV,  for the timelike scenario, is presented. Let us mention that it is also being addressed the running of the quark masses~\cite{Herren:2017osy}. Here, the decoupling nature of $\hat{\mu}_q$ is again observed. By taking as starting point the $Z$ gauge boson mass and ending with 1 TeV, the CMDM intensity decreases at most by 3 orders of magnitude, for instance, the most intense gap occurs for the up quark. In this analysis, which contemplates a timelike scenario and in contrast to the spacelike one, it should be noted the arising of imaginary contributions for each of the quarks under study. In the literature we have no references to compare our results except in the case of the charm quark. By fixing our predictions just at M=1000 GeV, the real and imaginary parts are $10^{-7}$ and $10^{-8}$, respectively, while the authors of the Ref. \cite{Hernandez-Juarez:2020drn} report $10^{-7}$ for both parts. Once again, we believe that this discrepancy is attributable to the fact that the authors of Ref. \cite{Hernandez-Juarez:2020drn} did not consider the running of the quark masses to the different energy scales of interest.

\section{Conclusions}
\label{Sec:conclusions}

We have calculated the anomalous chromomagnetic dipole moment for the SM quarks ($u$, $d$, $s$, $c$ and $b$) at the reference scale of the $Z$ gauge boson mass. In specific, we evaluated the off-shell gluon momentum transfer at $q^2=\pm m_Z^2$, since it is already known that in the SM an on-shell CMDM cannot be established due to the existence of an IR divergence, which comes from the non-Abelian triple gluon vertex. Additionally, for consistency, we have also considered the running of the quark masses, according to the $\overline{\textrm{MS}}$ scheme on that same energy scale. Our numerical results disagree with the literature, finding a discrepancy up to two orders of magnitude with previous reports.

\appendix
\section{Input values}
\label{appendix-Input-values}

We use the electron unit charge $e=\sqrt{4\pi\alpha}$ and the QCD group strong coupling constant $g_s=\sqrt{4\pi\alpha_s}$.
Our used input values, from PDG 2020 \cite{PDG2020}, are: the strong coupling constant $\alpha_s(m_Z)=0.1179$, the weak-mixing angle $s_W\equiv$ $\sin{\theta_W}(m_Z)$=$\sqrt{0.23121}$,
the boson masses $m_W$=$80.379$\;\;\; GeV, $m_Z$=$91.1876$ GeV, $m_H=125.1$ GeV and $m_t = 172.76$ GeV. The fine-structure constant $\alpha(m_Z)$ $=$ $1/129$ is taken from \cite{Denner:2019vbn}.

Regarding to the light quark masses, the provided values from PDG 2020 are given in the $\overline{\mathrm{MS}}$ scheme at low energy scale:
\begin{eqnarray}
m_u(2\thinspace\mathrm{GeV}) &=& 0.00216~ \mathrm{GeV},
\nonumber\\
m_d(2\thinspace\mathrm{GeV}) &=& 0.00467~ \mathrm{GeV},
\nonumber\\
m_s(2\thinspace\mathrm{GeV}) &=& 0.093~ \mathrm{GeV},
\nonumber\\
m_c(m_c) &=& 1.27~ \mathrm{GeV},
\nonumber\\
m_b(m_b) &=& 4.18~ \mathrm{GeV}.
\end{eqnarray}
Nevertheless, in order to evaluate in a consistent way the CMDMs at the scale of the $Z$ gauge boson mass, we obtain these running light quark masses at such scale through $\texttt{RunDec}$ \cite{Chetyrkin:2000yt,Herren:2017osy}, this by considering five active fermions and an accuracy of five loops:
\begin{eqnarray}
m_u(m_Z) &=& 0.00123~ \mathrm{GeV},
\nonumber\\
m_d(m_Z) &=& 0.00266~ \mathrm{GeV},
\nonumber\\
m_s(m_Z) &=& 0.05298~ \mathrm{GeV},
\nonumber\\
m_c(m_Z) &=& 0.6194~ \mathrm{GeV},
\nonumber\\
m_b(m_Z) &=& 2.874~ \mathrm{GeV}.
\end{eqnarray}

The quark-mixing matrix of Cabibbo-Kobayashi-Maskawa (CKM) \cite{PDG2020} is
\begin{eqnarray}\label{}
V_\mathrm{CKM}&=&
\left(
\begin{array}{ccc}
|V_{ud}| & |V_{us}| & |V_{ub}| \\
|V_{cd}| & |V_{cs}| & |V_{cb}| \\
|V_{td}| & |V_{ts}| & |V_{tb}| \\
\end{array}
\right)\nonumber\\
&=&\left(
\begin{array}{ccc}
 0.9737 & 0.2245 & 0.00382 \\
 0.221 & 0.987 & 0.041 \\
 0.008 & 0.0388 & 1.013 \\
\end{array}
\right).
\end{eqnarray}
The electric charges of the quarks are $Q_{u_i}=2/3$, $Q_{d_i}=-1/3$, and the weak charges $g_{Vu_i}=(3-8s_W^2)/6$, $g_{Au_i}=1/2$, $g_{Vd_i}=-(3-4s_W^2)/6$, $g_{Ad_i}=-1/2$.

\section{The Passarino-Veltman scalar functions}
\label{appendix-PaVe}

\begin{eqnarray}\label{C0-2}
&& C_0\left(m_q^2,m_q^2,q^2,m_q^2,m_X^2,m_q^2\right) \simeq
\nonumber\\
&&
\frac{1}{(q^2)^2}
\left\{
\left(2m_q^2+q^2\right)\left[
\mathrm{Li}_2\left(\frac{-2b}{a-2b+q^2}+q^2i\varepsilon\right)
\right.\right.
\nonumber\\
&&\left.\left.
+\mathrm{Li}_2\left(\frac{2b}{a+2b-q^2}+q^2i\varepsilon\right)
-\mathrm{Li}_2\left(\frac{2 (b-q^2)}{a+2b-q^2}+q^2i\varepsilon\right)
\right.\right.
\nonumber\\
&& \left.\left.
-\mathrm{Li}_2\frac{-2(b-q^2)}{a-2 b+q^2}
-\mathrm{Li}_2 \frac{b^2}{b^2+m_X^2q^2}
+\mathrm{Li}_2 \frac{b (b-q^2)}{b^2+m_X^2q^2}\right]
\right.
\nonumber\\
&& \left.
-\frac{2abm_q^2}{b^2+m_X^2q^2}\ln\frac{a+2m_q^2-q^2}{2 m_q^2}
-\frac{2m_q^2q^2}{b}
\right.
\nonumber\\
&& \left.
-\frac{2m_q^2m_X^2q^2\left(b^2-m_q^2q^2\right)}{b^2\left(b^2+m_X^2q^2\right)}\ln\frac{m_q^2}{m_X^2}
\right\},
\end{eqnarray}
with $a\equiv\sqrt{q^2(q^2-4m_q^2)}$ and $b\equiv m_q^2-m_X^2$. Here, $X$ can represent a Higgs boson or a $Z$ gauge boson.

\begin{eqnarray}\label{C0-5}
&& C_0(m_{q_i}^2,m_{q_i}^2,q^2,m_{q_j}^2,m_W^2,m_{q_j}^2) \simeq
\nonumber\\
&& \frac{1}{2q^2}
\Bigg\{2\left(2 m_{q_i}^2+q^2\right) \left[\mathrm{Li}_2\left(\frac{m_W^2}{q^2-m_{q_j}^2}+1\right)
\right.
\nonumber\\
&& \left.
+\mathrm{Li}_2\left(1-\frac{m_{q_j}^2}{m_W^2}\right)\right]
+\frac{2 m_{q_i}^2 q^2 \left(m_{q_j}^2-2 m_W^2\right)}{m_W^2 (m_{q_j}^2-m_W^2)}
\nonumber\\
&& +\left(2 m_{q_i}^2+q^2\right) \ln^2\frac{m_W^2}{m_{q_j}^2-q^2}\nonumber\\
&&-\frac{2 m_{q_i}^2 \left(m_{q_j}^2-2 m_W^2\right) \left(m_{q_j}^2-q^2\right)}{m_W^2 \left(m_{q_j}^2-m_W^2-q^2\right)}
\ln\frac{m_{q_j}^2}{m_{q_j}^2-q^2}
\nonumber\\
&& -\frac{2 m_{q_i}^2 q^2 \left[m_{q_j}^2 q^2-2 \left(m_{q_j}^2-m_W^2\right)^2\right]}{\left(m_{q_j}^2-m_W^2\right)^2 \left(m_{q_j}^2-m_W^2-q^2\right)}\ln \frac{m_{q_j}^2}{m_W^2}\Bigg\}.
\end{eqnarray}

\acknowledgments

This work has been partially supported by SNI-CONACYT and CIC-UMSNH. J. M. thanks C\'atedras CONACYT project 1753.

\end{document}